\documentclass[preprint,12pt,review]{elsarticle}

\usepackage{amssymb}

\journal{Physica A}

\begin{document}

\begin{frontmatter}



\title{Evidence of $q$-exponential statistics in Greek seismicity}


\address[label1]{Department of Physics, Institute for Complex Systems and Mathematical Biology (ICSMB), University of Aberdeen, AB24 3UE, Aberdeen, UK}
\address[label2]{Institute for Risk and Disaster Reduction, University College London, Gower Street, WC1E 6BT, London, UK}
\address[label3]{Laboratory of Geophysics and Seismology, Technological Educational Institute of Crete, Chania, GR 73133, Crete, Greece}
\address[label4]{Department of Mathematics, Center for Research and Applications of Nonlinear Systems (CRANS), University of Patras, GR 26110, Patras, Rio, Greece}

\cortext[cor1]{Corresponding author}
 
\author[label1]{Chris G. Antonopoulos\corref{cor1}}\ead{chris.antonopoulos@abdn.ac.uk}
\author[label2]{George Michas}\ead{georgios.michas.10@ucl.ac.uk}
\author[label2,label3]{Filippos Vallianatos}\ead{fvallian@chania.teicrete.gr}
\author[label4]{Tassos Bountis}\ead{bountis@math.upatras.gr}

\begin{abstract}
We study the seismicity (global seismic activity) that occurred in Greece between 1976 and 2009 based on the dataset reported in \cite{Makropoulosetal2012}, using concepts of Non-extensive Statistical Physics. By considering the entire and declustered datasets, for which the aftershocks have been removed, we initially investigate the frequency-magnitude distribution and find that both datasets are well approximated by a physical model derived in the framework of Non-extensive Statistical Physics. We then carry out a study of the distribution of interevent times of seismic events for different magnitude thresholds and discover that the data are well approximated by a statistical distribution of the $q$-exponential type that allows us to compute analytically the hazard function of earthquake production. Our analysis thus reveals further evidence that the underlying dynamical process of earthquake birth reflects a kind of nonlinear memory due to long-term persistence of seismic events.
\end{abstract}


\end{frontmatter}


\section{Introduction}

The Earth's crust can be considered as a complex dynamical system that interacts on a wide range of space and time scales to produce earthquakes, due to the relative motion of the tectonic plates. Typically, the time, location and magnitude of earthquakes are recorded to produce seismic catalogs that are further analyzed to study the physical patterns of seismicity. Understanding these patterns and the physical mechanism of the earthquake generation process still remains one of the main goals in Geophysics. Despite the complexity that is revealed through the analysis of seismic data, simple empirical relationships such as the Gutenberg-Richter \cite{Gutenbergetal1944} and the Omori law \cite{Omori1894} have long been recognized in earthquake sequences as indicating some kind of self-similarity and fractality in the earthquake generation process \cite{Turcotte1997}. The latter may be due to long-range correlations both in time and magnitude of the earthquakes (see e.g. \cite{Lennartzetal2008,Lennartzetal2011}).

In the present paper we study the earthquake activity in the geographical area of Greece in the time interval between 1976 and 2009 considering the larger earthquake magnitudes that occurred during this period. Taking into consideration the complex properties that are revealed in seismic catalogs, we use Non-extensive Statistical Physics (NESP) to study the frequency-magnitude and interevent time distributions. NESP has been introduced by Tsallis \cite{Tsallis1988} to propose a non-additive generalization of the classic Boltzmann-Gibbs entropy $S_{\mbox{BG}}$, hence allowing all-length scale correlations to interact within a given system. NESP has been successfully applied to a wide range of non-linear physical, social and artificial systems (see \cite{Tsallis2009}), where long-range correlations are leading to asymptotic power-law behavior in their statistical distributions. In earthquake physics it has been shown in a series of recent publications (e.g. \cite{Abeetal2003,Abeetal2005,Sotolongo-Costaetal2004,Vallianatosetal2013A,Papadakisetal2013,Michasetal2013} and references therein) that the statistical distributions that describe the size and spatio-temporal properties of seismicity can be related to the maximum entropy principle of the non-additive Tsallis entropy $S_q$.

Following this principle, we study the frequency-magnitude and the interevent time distributions of seismicity in Greece. In the first case, we use a physical model introduced in \cite{Sotolongo-Costaetal2004} and later revised in \cite{Silvaetal2006,Telesca2012}. This model considers that the released energy from the breakage of a fault is proportional to the volume of the fragments and asperities that fill the space between the fault planes. It has been recently applied to various local \cite{Vallianatosetal2013A,Michasetal2013} and regional earthquake data \cite{Telesca2010,Papadakisetal2013} and as our results indicate as well, it can be successfully applied to the seismicity in the area of Greece. Then, we also study the interevent time distribution $P_M(T)$ for various threshold magnitudes $M$ and find that in all cases it can be well approximated by a $q$-exponential function, known from NESP \cite{Tsallis2009}. The latter enables us to further evaluate analytically and present the hazard function $W_M(T,\Delta T)$ defined as the probability that at least one earthquake with magnitude larger than $M$ will occur in the next time interval $\Delta t$ if the last earthquake occurred $T$ days ago.

Our paper is organized as follows: in Sec. \ref{section_Analysis_of_Greek_Seismicity}, we analyze the seismological catalog for the area of Greece. Section \ref{section-The_hazard_function} is devoted to the analytical derivation of the hazard function and in Sec. \ref{section-Relation_between_R_M_and_M}, we report on a rough, approximated relation between the mean interevent time $R_M$ and magnitude $M$ of earthquake events that took place in the area of Greece. Finally, in the last section, we discuss our results and present the conclusions of our work.

\section{Analysis of Greek seismicity}\label{section_Analysis_of_Greek_Seismicity}

In this study we consider a recent catalog of seismicity reported in \cite{Makropoulosetal2012} spanning the period from 1901 to 2009. We have analyzed the shallow seismicity (i.e. for depths $\le40$km) that took place all over Greece, in the area confined to latitudes 34$^{o}$N to 42$^{o}$N and longitudes 19$^{o}$E to 29$^{o}$E for the period 1976-2009, as for this period the catalog can be considered complete for magnitudes $M$ greater than 4.1 (see \cite{Makropoulosetal2012}). Thus, in the analysis of the complete dataset which comprised 3523 events, we set throughout the paper the threshold magnitude to be $M_c=4.1$.

Further on, we have used the window method of \cite{Gardneretal1974}, as was later modified in \cite{Uhrhammeretal1986}, to identify the main shocks from the aftershocks and decluster the catalog keeping only the main seismic events. After the declustering procedure and for magnitudes $M\ge M_c$, a dataset of 2153 earthquakes remains. In both datasets we consider in our study, the maximum magnitude recorded is 6.9. The declustered dataset gives us the option to apply the analysis directly to the main earthquake events, in particular for the interevent times and the estimated hazard function. These results can then be compared to those obtained for the entire dataset to see how the aftershocks produced directly from the main shocks can influence the interevent time distribution, the estimated hazard function and the relation between index $q$ of Eq. (\ref{P_MT}), magnitude $M$ and mean interevent time $R_M$.

\subsection{Analysis of the frequency-magnitude distribution}\label{subsection-Analysis_of_the_frequency-magnitude_distribution}

In 2004, Sotolongo-Costa and Posadas starting from first principles developed a general physical model for the earthquake generation mechanism. In this model, the local breakage and the displacement of the asperities and fragments between the fault planes are the cause of the earthquake energy release. Accordingly, the released energy can be considered to be proportional to the volume of the fragments and the energy distribution function can be obtained in terms of the fragment size distribution \cite{Sotolongo-Costaetal2004}. These authors have considered that interactions between the fragments exist and are compatible with a model derived in the framework of NESP.

In terms of the probability $p(\sigma)$ of finding a fragment of surface size $\sigma$, Tsallis entropy $S_q$ is expressed as:
\begin{equation}
S_q=k_{\mbox{B}}\frac{1-\int p^q(\sigma)d\sigma}{q-1},
\end{equation}
where $k_{\mbox{B}}$ is Boltzmann's constant and $q$ is the so-called entropic index. For the sake of simplicity we set $k_{\mbox{B}}=1$. To find the probability $p(\sigma)$, the maximum entropy principle is applied under the appropriate constraints \cite{Tsallis2009}.

After the maximization procedure, the following expression for the fragment size distribution function is derived \cite{Silvaetal2006}:
\begin{equation}\label{P_sigma}
 P(\sigma)=\left[1-\biggl(\frac{1-q}{2-q}\biggr)(\sigma-\sigma_q)\right]^{\frac{1}{1-q}}.
\end{equation}
Assuming that the energy release $E$ is proportional to the volume of the fragments $E\sim r^3$ \cite{Silvaetal2006} and the magnitude $M$ is related to the energy $E$ as $M=\frac{2}{3}\log(E)$ \cite{Kanamori1978}, by integrating Eq. (\ref{P_sigma}), one can obtain the cumulative distribution (see \cite{Telesca2012,Michasetal2013}):
\begin{equation}\label{cumulative_distribution}
 \frac{N(>M)}{N}=\left[1-\biggl(\frac{1-q}{2-q}\biggr)\biggl(\frac{10^M}{a^{\frac{2}{3}}}\biggr)\right]^{\frac{2-q}{1-q}}.
\end{equation}
Eq. (\ref{cumulative_distribution}) describes from first principles, within the NESP formalism, the cumulative distribution of the number of earthquakes $N$ greater than the threshold magnitude $M$ (symbolized as $N(>M)$ herein) in a seismic region, normalized by the total number of earthquakes. Taking into account the minimum magnitude $M_0$ of the earthquake catalog that in our case is $M_0 = M_c$, the last equation should be slightly changed to \cite{Telesca2012}:
\begin{equation}\label{cumulative_distribution_complete_dataset}
 \frac{N(>M)}{N}=\left[\frac{1-\biggl(\frac{1-q}{2-q}\biggr)\biggl(\frac{10^M}{a^{\frac{2}{3}}}\biggr)}{1-\biggl(\frac{1-q}{2-q}\biggr)\biggl(\frac{10^{M_0}}{a^{\frac{2}{3}}}\biggr)}\right]^{\frac{2-q}{1-q}}.
\end{equation}

\begin{figure}[!ht]
\centering{
\includegraphics[scale=0.89]{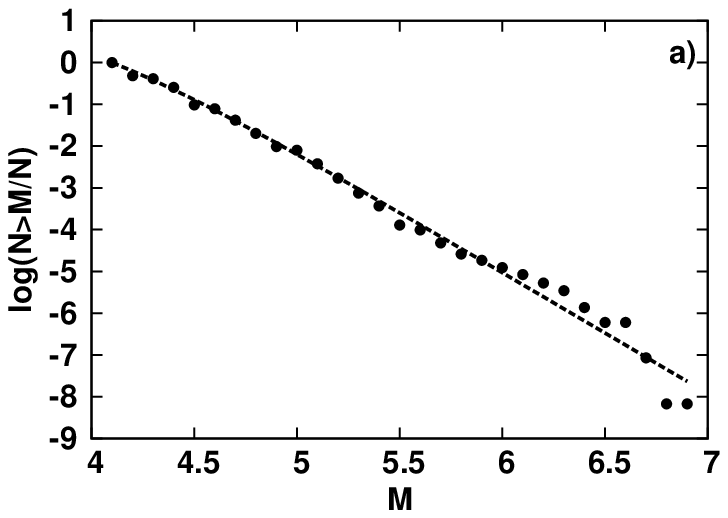}
\includegraphics[scale=0.89]{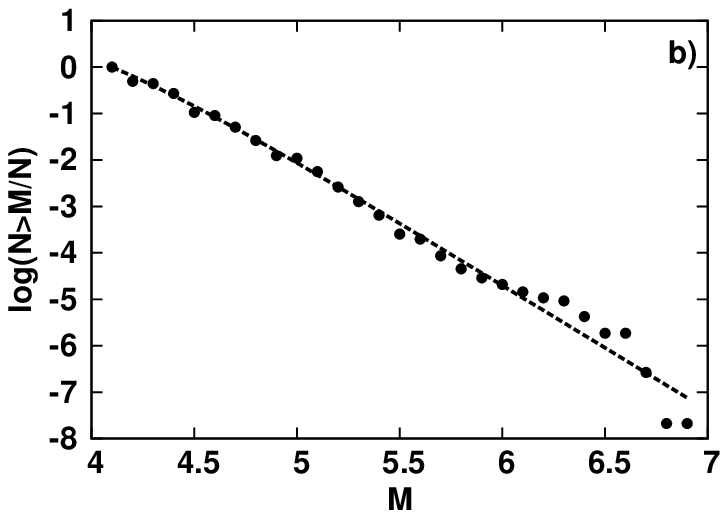}
}
\caption{Panel a): Normalized cumulative magnitude distribution (filled circles) for the entire dataset and for the model of Eq. (\ref{cumulative_distribution_complete_dataset}) (dashed line) for the values of $q_{1M}=1.443\pm0.018$ and $a_1=3.18\cdot10^5\pm1.7\cdot10^5$. Panel b): Normalized cumulative magnitude distribution (filled circles) for the declustered catalog and the model of Eq. (\ref{cumulative_distribution_complete_dataset}) (dashed line) for the values of $q_{2M}=1.46\pm0.018$ and $a_2=3.25\cdot10^5\pm1.7\cdot10^5$. In both panels, the dashed line has been obtained by performing a nonlinear fit to the data in black filled circles by the model function of Eq. (\ref{cumulative_distribution_complete_dataset}).}\label{fig1}
\end{figure}

We have applied this model to the earthquake magnitudes recorded in the area of Greece. Initially, we apply the model to the entire dataset and then to the declustered one. For the entire dataset the model describes quite well the observed distribution for the values of $q_{1M}=1.443\pm0.018$ and $a_1=3.18\cdot10^5\pm1.7\cdot10^5$ as one can see in panel a) of Fig. \ref{fig1}, while for the declustered one it describes it for the values of $q_{2M}=1.46\pm0.018$ and $a_2=3.25\cdot10^5\pm1.7\cdot10^5$ (see Fig. 1b)). In both cases, the values of $q$ and $a$ are quite similar indicating that the aftershocks included in the entire dataset do not alter significantly the observed cumulative frequency-magnitude distribution in the area of Greece for the period between 1976 and 2009.

\subsection{Analysis of the interevent time distribution}\label{subsection-Analysis_of_the_interevent_time_distribution}

The time evolution of seismic events in a geographical region is characterized by the set of discrete interevent times $T_i$ between seismic events $i=1,\ldots,n$ occurred in this region. For each magnitude threshold $M\ge M_c$ considered in our analysis, we study the distribution function $P_M(T)$ of the corresponding interevent times $T$. Since we deal with a finite and discrete number of recorded earthquake events and magnitudes (i.e. $n$ is finite), we compute the mean interevent time $R_M$ of the corresponding interevent times $T$ numerically as the mean of the interevent times considered for the estimation of each particular probability distribution $P_M(T)$. In practice, throughout the paper, we let $M$ range in $[4.1,5]$ and consider earthquake magnitudes between $M$ and the maximum magnitude $6.9$ of both datasets. We do that so that we are always left with enough data to produce reliable statistics for the calculation of $P_M(T)$.

The distribution $P_M(T)$ can be well approximated by a function of the form:
\begin{equation}\label{P_MT}
P_M (T)=\frac{A}{[1+(q-1)\beta T]^{\frac{1}{q-1}}},
\end{equation}
where $A$ is a normalization constant, and parameters $\beta,q$ depend on the fixed mean interevent time $R_M$. This form of $P_M(T)$ suggests that for $q>1$ interevent times of seismic events may be characterized by long-term memory effects related to correlation functions with power-law tails. The type of function appearing in Eq. (\ref{P_MT}) has the form of a ``generalized Pareto'' distribution \cite{Beirlantetal1996,Reissetal1997,Kleiberetal2003,Bercheretal2008} and is often called a $q$-exponential distribution. It is derived by maximizing $S_q$ under the appropriate constraints (see \cite{Tsallis2009}) and has been found to describe successfully the interevent time distribution of earthquake data for a variety of scales, from the laboratory to local, regional and global scales (e.g. \cite{Abeetal2005,Vallianatosetal2012,Vallianatosetal2013B,Vallianatosetal2013A,Papadakisetal2013,Michasetal2013}). If $T$ is replaced by $T^2$ in the denominator of Eq. (\ref{P_MT}), the latter equation becomes the $q$-Gaussian probability distribution function known from Non-extensive Statistical Mechanics \cite{Tsallis2009,Bountisetal2012}.

\begin{figure}[!ht]
\centering{
\includegraphics[scale=0.89]{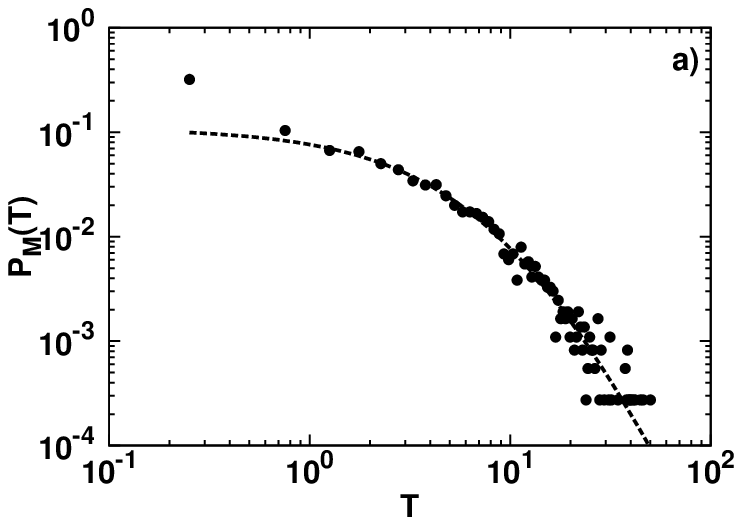}
\includegraphics[scale=0.89]{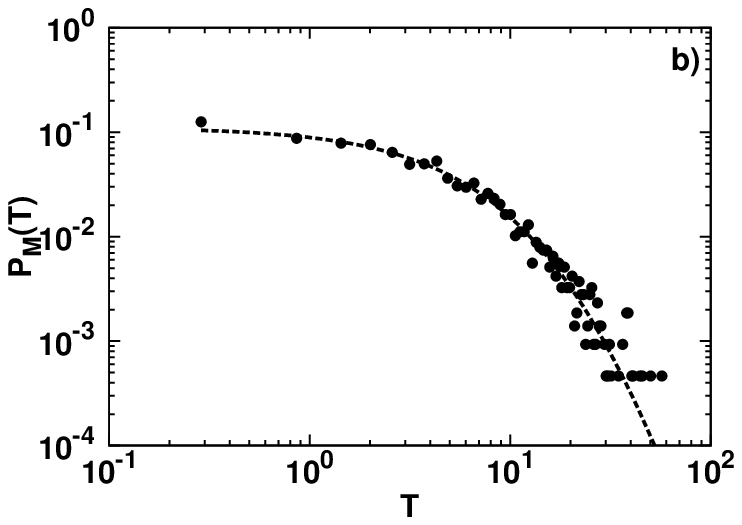}
}
\caption{Panel a): Plot of the interevent time distribution $P_M(T)$ versus the interevent time $T$ (in days) for earthquake magnitudes $M\ge M_c$ for the entire dataset considered. The dashed curve is the fit of the data (in filled circles) by the $q$-exponential function of Eq. (\ref{P_MT}). Here we have $q_{T1}=1.24\pm0.054$. Panel b): Same as in a) but for the corresponding declustered dataset. In this case we obtain $q_{T2}=1.14\pm0.057$. Note that all axes are logarithmic.}\label{fig2}
\end{figure}

In Fig. \ref{fig2}a) we demonstrate a particular example of the interevent time distribution $P_M(T)$ versus the interevent time $T$ (expressed in days) for magnitudes $M\ge M_c$ for the entire dataset. The dashed curve is the numerical fit of the data (filled circles) by the $q$-exponential function of Eq. (\ref{P_MT}), giving $q_{T1}=1.24\pm0.054$. Fig. \ref{fig2}b) presents the $q$-exponential fitting of the interevent times $T$ of the declustered dataset for $M$ values in the same magnitude interval. Here, the $q$-exponential fitting leads to $q_{T2}=1.14\pm0.057$, a smaller value than the one of panel a), indicating that the included aftershocks in the first case have the tendency to increase the $q$-value in the interevent time distribution, probably due to clustering effects observed in aftershock sequences that lead in larger deviations from the pure exponential function recovered from Eq. (\ref{P_MT}) in the limit $q\rightarrow1$.

Our results show that index $q$ in Eq. (\ref{P_MT}) attains values bigger than $q=1$ of the simple exponential distribution $e^{-\beta T}$ in agreement with \cite{Abeetal2005}, in the case of long term complete datasets that include main events and aftershocks. This result also suggests that main earthquakes accompanied by their aftershocks are more strongly time-correlated and hence lie further away from purely exponential statistics, which govern strongly chaotic processes with $q\rightarrow1$ \cite{Bountisetal2012}.

\section{The hazard function}\label{section-The_hazard_function}

Having thus an analytical expression available for the distribution $P_M(T)$, makes it interesting to study the hazard function $W_M(T,\Delta T)$ defined as the probability that at least one earthquake with magnitude bigger than $M_c$ will occur in the next time interval $\Delta T$ if the last earthquake occurred $T$ days ago. In particular, $W_M$ and $P_M$ are related by \cite{Ludescheretal2011}:
\begin{equation}\label{hazard_function_initial}
W_M(T,\Delta T)=\frac{\int_T^{T+\Delta T} P_M(t)dt}{\int_T^{\infty}P_M(t)dt}.
\end{equation}
By substituting $P_M(T)$ from Eq. (\ref{P_MT}) into Eq. (\ref{hazard_function_initial}), one easily derives by direct integration:
\begin{equation}\label{hazard_function_final}
W_M(T,\Delta T)=1-\left[1+\frac{\beta(q-1)\Delta T}{1+\beta(q-1)T}\right]^{\frac{q-2}{q-1}}.
\end{equation}

\begin{figure}[!ht]
\centering{
\includegraphics[scale=0.89]{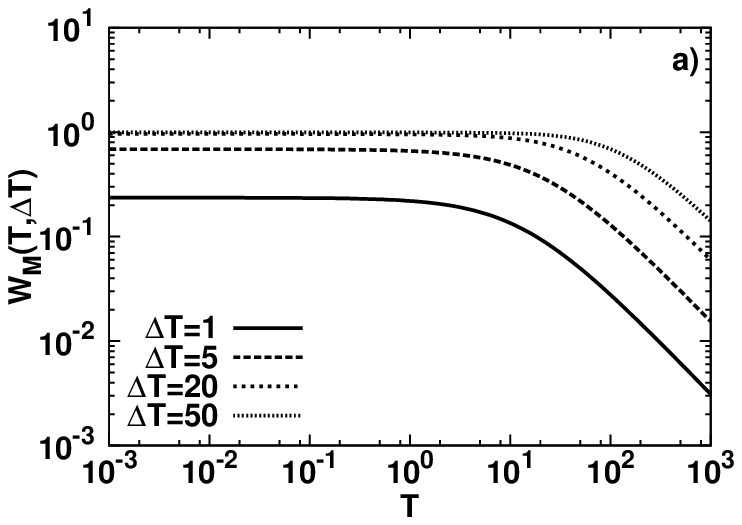}
\includegraphics[scale=0.89]{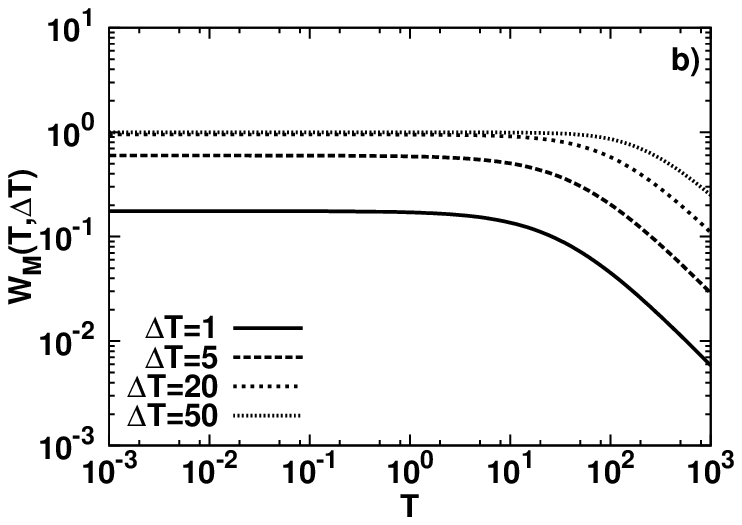}
}
\caption{Panel a): Plot of the hazard function $W_M(T,\Delta T)$ of Eq. (\ref{hazard_function_final}) versus the interevent time $T$ (in days) for earthquake magnitude thresholds $M\ge M_c$ for the entire dataset considered in this paper. We have used four different $\Delta T$s and the same values for the parameters $\beta,q$ as in panel a) of Fig. \ref{fig2}. Panel b): Same as in a) for the corresponding declustered dataset and the values of Fig. \ref{fig2}b). Note that all axes are logarithmic.}\label{fig3}
\end{figure}

It is straightforward to prove that for exponentially decaying distributions $P_M(T)$, the hazard function $W_M(\Delta T)=1-e^{-\beta \Delta T}$ and is hence independent of the interevent time $T$, while for probability distribution functions decaying by a power law, $W_M(T,\Delta T)\propto \frac{\Delta T}{T}$ for $\Delta T\ll T$. We present such examples in Fig. \ref{fig3} where we plot the hazard function $W_M(T,\Delta T)$ of Eq. (\ref{hazard_function_final}) for four different interevent time intervals $\Delta T$ for the two datasets we have considered. We use in each case the datasets and $\beta,q,M$ values of the corresponding panels of Fig. \ref{fig2}. It is apparent for both datasets that, for a fixed time interval $\Delta T$, the probability that at least one earthquake with magnitude $M\ge M_c$ will occur in the next time interval $\Delta T$ (if the last earthquake occurred $T$ days ago) decreases as $T$ increases.

Moreover, if we increase $\Delta T$ the probability $W_M(T,\Delta T)$ increases as well. The curve of the hazard function consists of two parts: the left one which is related to the exponentially decaying part of the curve of the corresponding interevent time distribution and the right one which is associated with the power-law decay of the interevent time distribution (see Fig. \ref{fig2}).

\section{An approximate relation between $R_M$ and $M$}\label{section-Relation_between_R_M_and_M}

Finally, based on the results we have presented in the previous sections, we attempt here to establish an approximate functional dependence between $R_M$ and $M$ for the two datasets (i.e. entire and declustered) considered in this paper. We focus only on this relation since the error bars compatible with data reported in \cite{Makropoulosetal2012} on the dependence of $q$ on $M$ and $q$ on $R_M$ are too large to permit accurate estimates. Moreover, the relation between $R_M$ and $M$ for both datasets is particularly useful as it can serve as a rough predictor of the mean interevent time of seismic events in Greece for given earthquake magnitudes $M\ge M_c$.

\begin{figure}[!ht]
\centering{
\includegraphics[scale=0.58]{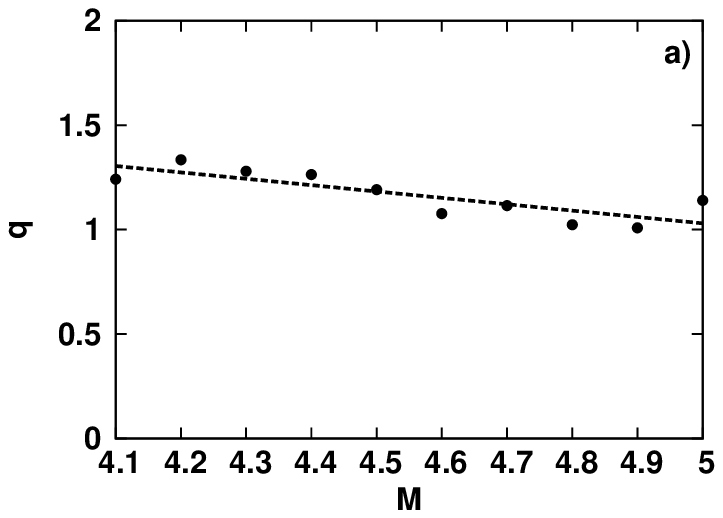}
\includegraphics[scale=0.58]{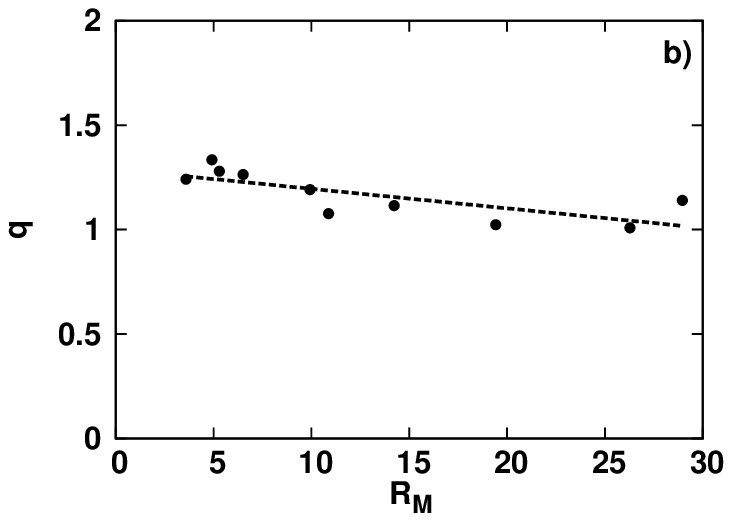}
\includegraphics[scale=0.58]{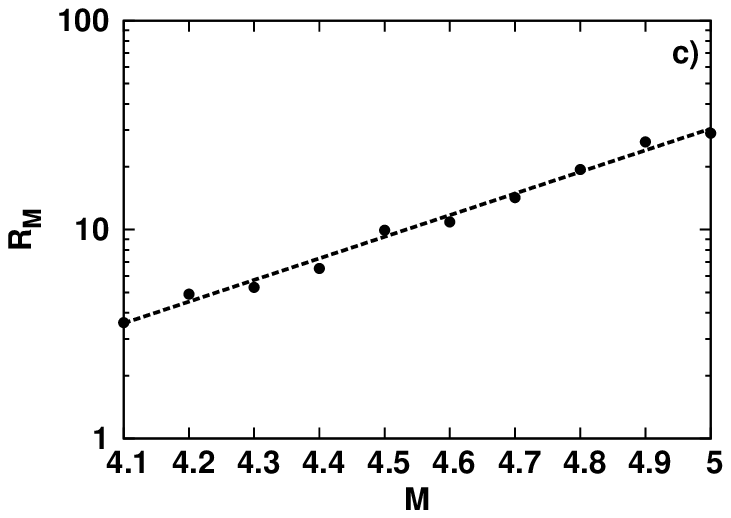}\\
\includegraphics[scale=0.58]{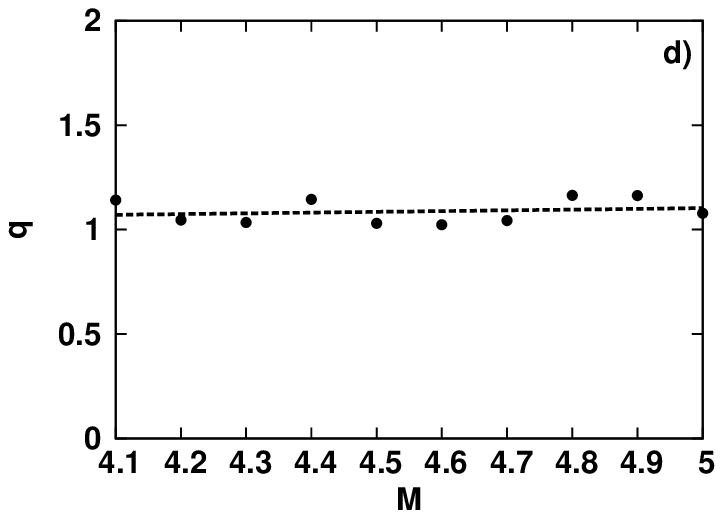}
\includegraphics[scale=0.58]{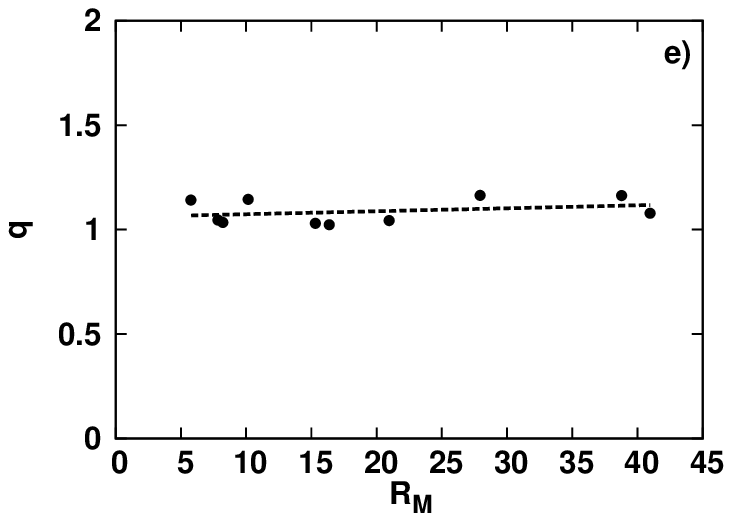}
\includegraphics[scale=0.58]{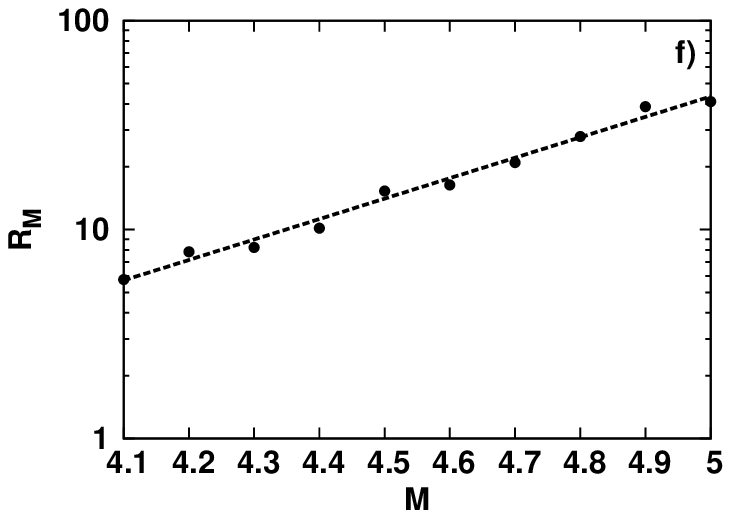}
}
\caption{Panel a): Plot of $q$ versus $M$ for earthquake magnitudes $M\ge M_c$ for the entire dataset considered in this paper. Panel b): Plot of $q$ versus $R_M$ for the same magnitudes and dataset as in panel a). Panel c): Plot of the mean interevent time $R_M$ versus the magnitude $M$ for the same magnitudes and dataset as in panel a). Panels d), e) and f): Same as in panels a), b) and c) for the corresponding declustered dataset. In panels a), b) and d), e) we also plot with dashed lines the linear trend of the data to guide the eye and in panels c) and f) the approximate fitting function of Eq. (\ref{RM_vs_M}) (dashed line) to the data (filled circles). Note that the vertical axes of panels c) and f) are logarithmic.}\label{fig4}
\end{figure}

Thus, in Fig. \ref{fig4} we plot $q$ versus $M$ in panel a), $q$ versus $R_M$ in panel b) and $R_M$ versus $M$ in panel c) for the entire dataset and for the corresponding declustered dataset in panels d), e) and f) respectively. In panels a), b) and d), e) we also plot with dashed lines the linear trends of the data to guide the eye. A comparison of these trends reveals that for the declustered dataset, $q$ remains almost constant as a function of $M$ and $R_M$ (see panels d) and e) respectively) while it decreases to 1 in the case of the entire dataset (see panels a) and b)).

Passing now to a possible relation between $R_M$ and $M$, we propose that:
\begin{equation}\label{RM_vs_M}
\log(R_M)=\log(a) + b M,
\end{equation}
since, as we demonstrate in panels c) and f) of Fig. \ref{fig4}, the points fall quite nicely to a line in linear-log plot. In more detail, we have been able to fit the data of both datasets shown in panels c) and f) of Fig. \ref{fig4} using Eq. (\ref{RM_vs_M}). For the entire dataset shown in panel c) we find $a=2.04\cdot 10^{-4}\pm1.22\cdot10^{-4}$ and $b=2.38\pm0.12$, while for the declustered one shown in panel f) we get $a=5.58\cdot10^{-4}\pm3.74\cdot10^{-4}$ and $b=2.25\pm0.14$. We find that the values of the exponents $b$ of the two datasets are quite close indicating an almost similar trend of $R_M$ to increase monotonically as $M$ increases. Equations such as (\ref{RM_vs_M}) can serve as a rough predictor of the mean interevent time of seismic events in Greece for given earthquake magnitudes $M\ge4.1$.

These findings allow us to argue that as the magnitude $M$ and mean interevent time $R_M$ of seismic events increase when one considers the entire dataset, $q$ starts from values higher than 1 and gradually approaches 1 of Gaussian distributions, meaning that the dynamics responsible for the generation of earthquakes becomes strongly chaotic. In contrast, when one considers the declustered dataset with the aftershocks removed, $q$ attains values already very close to 1 even for the relatively smaller $M$ and $R_M$ values. These findings indicate that aftershocks are responsible for the increase of $q$ that further supports the conjecture that the underlying dynamical process reflects a kind of nonlinear memory due to long-term persistence effects \cite{Lennartzetal2008,Lennartzetal2011}. This leads to the conclusion that interevent times are long-term correlated and possess autocorrelation functions that decay by power-law.


\section{Conclusions}

In this paper we have studied the seismicity that occurred in the geographical area of Greece using concepts from Non-extensive Statistical Physics based on the dataset reported in \cite{Makropoulosetal2012}. We have considered in our study the entire dataset for $M\ge4.1$ for the period 1976-2009, as well as the corresponding declustered dataset, for which the aftershocks have been removed. Initially, we explored the frequency-magnitude distribution and found that both datasets can be well approximated by a physical model derived in the NESP framework and for similar values of the fitting parameters. The values of $q$ and $a$ thus estimated can be used to reproduce the size distribution of earthquakes in Greece for the considered period.

Next, we studied the distribution of interevent times $T$ for different magnitude thresholds and found for both datasets, that the data are well approximated by a statistical distribution of the $q$-exponential form shown in  Eq. (\ref{P_MT}). The form of this distribution enabled us to compute analytically the hazard function representing the probability that at least one earthquake of magnitude larger than $M$ will occur in the next time interval $\Delta T$, if the last earthquake occurred $T$ days ago. We have thus obtained, for both datasets and for a fixed time interval $\Delta T$, the probability that at least one earthquake event with magnitude $M\ge4.1$ will occur in the next time interval $\Delta T$ (if the last earthquake occurred $T$ days ago) decreases as $T$ increases and that, if we increase $\Delta T$, the probability $W_M(T,\Delta T)$ increases as well.

Finally, we presented an approximate, roughly estimated functional relation between $R_M$ and $M$ that can serve as a rough predictor of the mean interevent time of seismic events in Greece for given earthquake magnitudes $M\ge4.1$. Our analysis has revealed further evidence that aftershocks of main seismic events are responsible for the increase of the $q$ index that we believe supports further the conjecture that the underlying dynamical process of earthquake generation reflects a kind of nonlinear memory due to long-term persistence effects, and thus leading to the conclusion that interevent times are long-term correlated and possess autocorrelation functions that decay by power-law.


\section{Acknowledgments} 

The research of C.G. A. and T. B. has been co-financed by the European Union (European Social Fund--ESF) and Greek national funds through the MACOMSYS Project ``Mathematical Modeling of Complex Systems'', within the Operational Program ¡¡Education and Lifelong Learning¢¢ of the National Strategic Reference Framework (NSRF)-Research Funding Program: THALES-Investing in knowledge society through the European Social Fund (MIS 379337). The work of F. V. and G. M. was supported by the THALES program of the Ministry of Education of Greece and the European Union in the framework of the project entitled ``Integrated understanding of Seismicity, using innovative Methodologies of Fracture mechanics (MIS 380208) along with Earthquake and Non-extensive statistical physics - Application to the geodynamic system of the Hellenic Arc. SEISMO FEAR HELLARC''.




$\bibliographystyle{elsarticle-num}
$\bibliography{bibliography}


\end{document}